\documentclass[twocolumn,twoside,showkeys,showpacs,preprintnumbers,superscriptaddress,amsmath,amssymb,prc]{revtex4-2}

\usepackage{graphicx}
\usepackage{amsmath}    
\usepackage{amssymb}    
\usepackage{dcolumn}
\usepackage{bm}
\usepackage[version=4]{mhchem}
\usepackage{appendix}

\usepackage{float}
\usepackage{subcaption}
\usepackage[justification=raggedright,singlelinecheck=false]{caption}

\usepackage{tikz}
\usepackage{ulem}
\usepackage{changes}

\usepackage[colorlinks=true,linkcolor=black,citecolor=blue,urlcolor=blue]{hyperref}

\usepackage[T1]{fontenc}

\begin{document}

\title{Probing the Three-dimension Emission Source and Neutron Skin via $\pi$-$\pi$ Correlations in Heavy-Ion Collisions}

\author{Haojie Zhang}
\email{zhang-hj22@mails.tsinghua.edu.cn}
\affiliation{Department of Physics, Tsinghua University, Beijing 100084, China}

\author{Junhuai Xu}
\affiliation{Department of Physics, Tsinghua University, Beijing 100084, China}

\author{Pengcheng Li}
\affiliation{School of Science, Huzhou University, Huzhou 313000, China}

\author{Zhi Qin}
\affiliation{Department of Physics, Tsinghua University, Beijing 100084, China}

\author{Dawei Si}
\affiliation{Department of Physics, Tsinghua University, Beijing 100084, China}

\author{Yijie Wang}
\affiliation{Department of Physics, Tsinghua University, Beijing 100084, China}

\author{Yongjia Wang}
\affiliation{School of Science, Huzhou University, Huzhou 313000, China}

\author{Qingfeng Li}
\email{liqf@zjhu.edu.cn}
\affiliation{School of Science, Huzhou University, Huzhou 313000, China}


\author{Zhigang Xiao}
\email{xiaozg@tsinghua.edu.cn}
\affiliation{Department of Physics, Tsinghua University, Beijing 100084, China}
\affiliation{Center of High Energy Physics, Tsinghua University, Beijing 100084, China}

\date{\today}

\begin{abstract}


The Richardson–Lucy algorithm is applied to reconstruct the three-dimensional source function of identical pions from their two-particle correlation functions. The algorithm's performance is first evaluated through simulations with Gaussian-type initial source functions. Its imaging quality and robustness are further demonstrated with experimental data from Au+Au collisions at 1.23 A GeV, collected by the HADES Collaboration. Additionally, using UrQMD simulations of Pb+Pb collisions at 1.5 A GeV, we show that the deblurred source functions exhibit sensitivity to the initial neutron skin thickness of the colliding nuclei. This highlights the potential of the Richardson–Lucy algorithm as a tool for probing the neutron density distribution in heavy nuclei.

\end{abstract}

\maketitle

\section{Introduction}

Two-particle femtoscopy, commonly referred to as Hanbury Brown–Twiss (HBT) interferometry, was first introduced by Hanbury Brown and Twiss \cite{HanburyBrown:1956bqd} to determine the angular radii of stars and has since been enormously employed in femtoscopic studies of nuclear collisions over wide energy range \cite{Goldhaber:1960sf, Lisa:2005dd,Ghetti:2003zz,PhysRevLett.67.14,Wang:2021mrv,PhysRevLett.134.222301,PhysRevC.93.024905,STAR:2015kha,PhysRevC.20.2267}. The importance of the correlation function stems from its connection to the source function, a key physical quantity that encodes the spatial geometry of the collision fireball at freeze-out. This connection is typically established using the Kopylov-Podgoretsky (KP) formalism \cite{Grishin:1971wu, Koonin:1977fh, Lisa:2005dd}, which relates the correlation function to the source function via the two-particle interaction wave function.

In practice, the source function is often approximated by a Gaussian form, with the corresponding source parameters, commonly referred to as HBT radii, obtained from fits to the correlation function. The source function offers valuable insights into the space-time evolution of the emitting source, revealing features such as the freeze-out hypersurface \cite{Heinz:2002un}, collective flow patterns \cite{Lisa:2005dd}, and the equation of state (EoS) \cite{HADES:2020ver}, particularly in the context of phase transitions between hardon gas and quark-gluon plasma \cite{Pratt:1986cc, Li:2022iil,Rischke:1996em}.  Therefore, accurate reconstruction of the source function remains a central objective in correlation function studies.

With the potential of extracting the spatial-temporal information of the particle emissions source, it is a natural attempt  to  probe the neutron skin  of the colliding nuclei using femtoscopy. Such motivation stems from the facts that the neutron skin thickness $\Delta R_{\rm np}$ of heavy nuclei provide valuable constraints on the density dependence of nuclear symmetry energy \cite{Brown:2000pd,PREX:2021umo,Reed:2021nqk,CREX:2022kgg,Reinhard:2022inh}, 
a focusing frontier of nuclear physics and astrophysics. It has been proposed to probe the neutron halo by measuring the neutron-neutron correlation function in reactions induced by neutron-rich nuclei \cite{Cao:2012cv}. 
However, after a violent collision, the  colliding nuclei are smashed and their initial static structures are destructed, making this goal extremely challenging. Enormous efforts are definitely required, both experimentally and theoretically,  to resume the initial neutron distribution from the freeze-out state. To this end, one needs not only transport model tracing the evolution of the nucleons in the collision, but also precise source imaging method to extract the fine information of the neutron distribution profile at freeze-out instant based on  high-statistics data. 


In this work, we employ the Richardson–Lucy (RL) algorithm, a model-independent approach rooted in Bayesian inference, to reconstruct the three-dimensional source functions from $\pi\pi$ pair correlation function \cite{Brown:2005ze}. While some prior studies have realized one-dimensional source imaging \cite{Xu:2024dnd, Tam:2025mkk}, we extend the methodology to three dimensions in order to capture a more complete picture of the spatial source structure. Using UrQMD simulation data, we further investigate how sensitive the neutron skin thickness affect the reconstructed source function.

Our choice to focus on $\pi\pi$ correlation functions is motivated by several factors:

\begin{itemize}
\item Pions are abundantly produced in heavy-ion collisions and can be measured with high precision, providing a robust experimental dataset.
\item The interactions between pions are dominated by the Coulomb interaction, which significantly simplifies the wave function treatment in the KP formalism.
\item The yields of charged pions are sensitive to the isospin composition of the colliding system, making them a potential probe of neutron distribution in the femtoscopic imaging application.
\end{itemize}

This paper is organized as follows. Section II outlines the theoretical background and the implementation of the Richardson–Lucy algorithm. In Section III, the method is validated through two complementary approaches: first, using simulated data from known Gaussian source functions, and second, applying it to experimental data from the HADES Collaboration in 2018 \cite{HADES:2018gop}, both serving to verify the effectiveness and reliability of the reconstruction. Section IV presents the application to UrQMD simulations, illustrating how the extracted source functions respond to variations in the neutron skin thickness. Finally, a summary and conclusions are provided in the last section.

\section{Theoretical Analysis}

\subsection{Correlation Function and Femtoscopy}

The correlation function $C(\mathbf{q})$ serves as an essential observable in heavy-ion collisions, providing insight into the spatial and temporal characteristics of the particle-emitting source by measuring the momentum correlations between two particles. Experimentally, it is constructed as the ratio of the distribution of particle pairs from the same event to that from an uncorrelated mixed-event background:

\begin{equation}
C(\mathbf{p_1,p_2}) \propto \frac{A_{\text{same}}\left(\mathbf{p_1}, \mathbf{p_2}\right)}{B_{\text{mix}}\left(\mathbf{p_1}, \mathbf{p_2}\right)},
\end{equation}
where $\mathbf{p_1}$ and $\mathbf{p_2}$ represent the momenta of two particles from the same event or from mixed events, respectively.  The function is normalized by the number of pairs in the mixed-event sample. It is often convenient to work in the center-of-mass frame, where the correlation function is expressed as $ C(\mathbf{q}, \mathbf{Q}) $, with $ \mathbf{q} = \mathbf{q}_1 - \mathbf{q}_2 $ being the relative momentum, and $ \mathbf{Q} = \mathbf{q}_1 + \mathbf{q}_2 $ is the total momentum.
In many analyses, the dependence on $ \mathbf{Q}$ is integrated out, reducing the correlation function to $ C(\mathbf{q}) $.

The shape of $ C(\mathbf{q}) $ reveals important physical effects: at low relative momentum, it deviates from unity due to final-state interactions such as Coulomb and strong forces, as well as quantum statistics; at large $\mathbf{q}$ where correlations vanish, it approaches unity.

In the KP model, the two-particle correlation function  is related to the underlying source distribution via the expression:

\begin{equation}
C(\mathbf{q}) = \int S(\mathbf{r}) \left| \psi(\mathbf{q}, \mathbf{r}) \right|^2 \mathrm{d}^3\mathbf{r},
\end{equation}

where  $S(\mathbf{r}) $ is the source function, representing the spatial distribution of particle emission points in the pair rest frame. It is important to note that although $S(\mathbf{r})$ depends only on the spatial coordinate $\mathbf{r}$, it implicitly corresponds to an average over momentum space, with the momentum distribution being determined experimentally for a given kinematic region.  $ \psi(\mathbf{q}, \mathbf{r})$ denotes the two-particle scattering wave function, which encodes final-state interactions
including Coulomb and strong forces. 

Experimentally, the relative momentum $ \mathbf{q} $  is commonly decomposed in the longitudinally co-moving system (LCMS) with respect to the average transverse momentum of the pair,
$ \mathbf{k}_{\rm T} = \frac{\mathbf{k}_{\rm T1} + \mathbf{k}_{\rm T2}}{2} $. Following the convention introduced by Podgoretsky, Pratt, and Bertsch \cite{Podgoretsky:1982xu,Pratt:1986cc,Bertsch:1989vn}, it is split into three orthogonal components: 
\begin{itemize}
\item $ q_{\text{out}} $: along  $ \mathbf{k}_{\rm T} $.
\item $ q_{\text{side}} $: perpendicular to $ q_{\text{out}} $ and the beam.
\item $q_{\text{long}} $: along the beam. 
\end{itemize}
In this basis, the correlation function becomes $ C(q_o, q_s, q_l) $, allowing for a three-dimensional HBT analysis of the source geometry. 

By fitting the measured correlation function to the KP formula, key properties of the emission source can be quantitatively inferred. A common approach is to assume that the source function $ S(\mathbf{r}) $ follows a three-dimensional Gaussian profile in the out-side-long (o-s-l) coordinate system:

\begin{equation}
S(r_o, r_s, r_l) \propto \exp[-\frac{1}{2}(\frac{r_o^2}{R_{o}^2} + \frac{r_s^2}{R_{s}^2} + \frac{r_l^2}{R_{l}^2})],
\end{equation} 
where $ R_{o} $, $ R_{s} $, and $ R_{l} $ denote the Gaussian source radii in the out, side, and long directions, respectively, and $ r_o, r_s, r_l $ are the corresponding spatial coordinates in the pair rest frame.

These radii are typically extracted by fitting the correlation function using the Bowler–Sinyukov method \cite{Sinyukov:1998fc}, which employs the functional form:

\begin{equation}
C(\mathbf{q}) = (1 - \lambda) + \lambda K(\mathbf{q}) \left[ 1 + \exp \left( -\sum_{i=o, s, l} q_i^2 R_i^2 \right) \right].
\end{equation}
Here, $\lambda$ represents the correlation strength parameter, quantifying the fraction of particles originating from chaotic emission, with $( 1-\lambda )$ interpreted as the coherent fraction.
The factor $ K(\mathbf{q}) $ corrects for Coulomb interactions between charged particles. It is worth noting that some recent studies also introduce cross terms, such as $ R_{o-l}$ to capture more complex source shapes \cite{Luong:2024eaq}.

This fitting framework offers a practical advantage: it avoids direct numerical integration of the full two-particle Coulomb wave function, significantly simplifying the extraction of source radii while maintaining physical consistency.

\subsection{Relative Scattering Wave Function of Pions}

To compute the correlation function for identical $\pi$ pairs, the two-particle wave function must be constructed with Coulomb final-state interactions taken into account. In the center-of-mass frame, the stationary Schrödinger equation for the relative motion reads:

\begin{equation}
    [ -\frac{\hbar^2}{2\mu} \nabla^2 + \frac{Z_1 Z_2 e^2}{r} ] \psi_{\rm C}(\mathbf{r}) = E \psi_{\rm C}(\mathbf{r}),
\end{equation}

where $\psi_{\rm C}(\mathbf{r})$ is the Coulomb wave function, $\mu = m_{\pi}/2$ is the reduced mass for two identical pions, and $Z_1=Z_2=\pm{1}$ correspond to $\pi^+\pi^+$ or $\pi^-\pi^-$ pairs. 

An analytic solution can be obtained using separation of variables in cylindrical coordinates. The resulting wave function takes the form:

\begin{equation}
    \psi(r, z) =  e^{i k z} F(-i \eta, 1, k(r - z)),
\end{equation}

where $ k=\sqrt{2\mu E}/\hbar $ is the relative wave number, $Ac(\eta) = \frac{2\pi \eta}{e^{2\pi \eta} - 1}$ is known as the Gamow factor, describing Coulomb penetration effects, and $\eta = \frac{\alpha Z_1 Z_2 c \mu}{\hbar k}$,  with $\alpha$ being the fine-structure constant. The term $F(-i \eta, 1, k(r - z))$ is the confluent hypergeometric function.

\begin{figure}[hptb]
    \centering
    \includegraphics[width=0.45\textwidth]{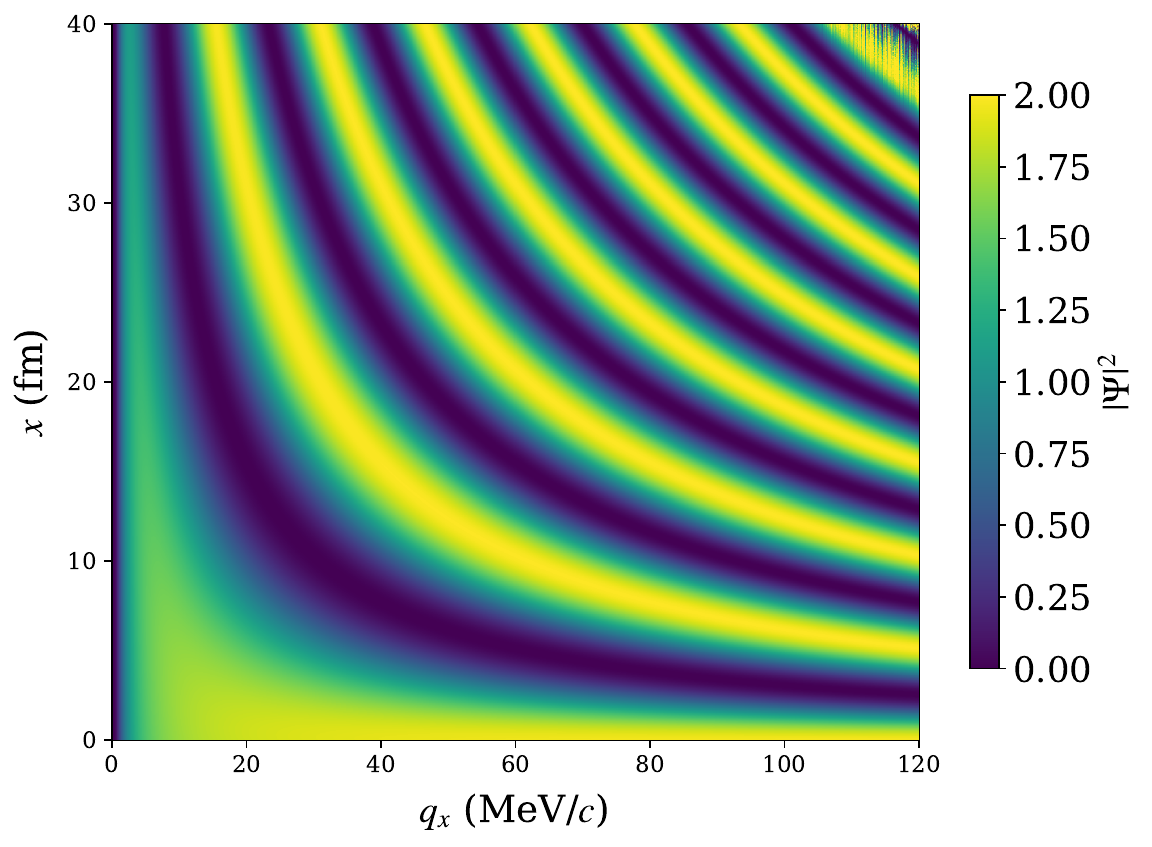}
    \caption{Coulomb wave function modulus square $|\Psi(x, q_x)|^2$ as a function of coordinate space position $x$ and momentum space component $q_x$. The color intensity represents the probability density, with brighter regions indicating higher values.}
    \label{fig1:wf}
\end{figure}

For identical bosons such as $\pi^+\pi^+$ or $\pi^-\pi^-$, which are scalar bosons with symmetric spin, the total wave function must be symmetric under particle exchange, as written below 
\begin{equation}
    \psi_{\pi^+\pi^+(\pi^-\pi^-)} = \frac{1}{\sqrt{2}} \left[ \psi(r) + \psi(-r) \right].
\end{equation}

This symmetrized form is used in KP formula for the $\pi\pi$ correlation function. As shown in Fig.~\ref{fig1:wf}, the Coulomb wave function distribution exhibits the expected behavior in the $x$-$q_x$ plane, with other degrees of freedom frozen. The computational divergence at large $x$ and $q_X$ values, visible in the upper right corner, stems from the limitations of the confluent hypergeometric function evaluation. However, this numerical issue does not affect the physical results since our analysis focuses on momentum differences typical of HBT radii below 20 fm, where the wave function remains well-behaved and accurately computable.

\subsection{Richardson-Lucy algorithm}

Originally developed for image restoration \cite{Patel2015}, the Richardson–Lucy (RL) algorithm is an iterative deconvolution technique widely employed in astronomy, microscopy, and signal processing to recover an original image from a blurred observation. Recently, it has found broad application in nuclear physics \cite{Danielewicz:2021vqq, Nzabahimana:2023tab, Xu:2024oct, Xu:2025luc}.  It operates within a maximum-likelihood framework, reconstructing the underlying distribution $\mathcal{F}(\mu) $ from a measured signal $ f(\nu) $ that has been convolved with a known point spread function (PSF) $ P(\nu \mid \mu)$: 

\begin{equation}
    f(\nu) = \int \mathrm{d}\mu P(\nu \mid \mu) \mathcal{F}(\mu),
\end{equation}where $ P(\nu \mid \mu) $ denotes the conditional probability of observing $\nu$ given the true value $\mu$. The RL algorithm iteratively improves the estimate of $\mathcal{F}(\mu) $ by comparing the current prediction $ f^{(r)}(\nu) $ with the actual measurement $ f(\nu) $. The update rule at each iteration is given by:

\begin{equation}
    \begin{aligned}
    \mathcal{F}^{(r+1)}(\mu) &= \mathcal{F}^{(r)}(\mu) \frac{\int d\nu \frac{f(\nu)}{f^{(r)}(\nu)} W(\nu) P(\nu \mid \mu)}{\int d\nu' W(\nu') P(\nu' \mid \mu)} \\
    &= A^{(r)}(\mu) \cdot \mathcal{F}^{(r)}(\mu),
    \end{aligned}
\end{equation}where $r$ is the iteration index and $ W(\nu) $ represents a relative weighting factor for the measured values. The predicted observation at the 
$r$-th iteration is computed as:

\begin{equation}
    f^{(r)}(\nu) = \int \mathrm{d}\mu P(\nu \mid \mu) \mathcal{F}^{(r)}(\mu).
\end{equation}

Convergence is monitored using a $\chi^2$ statistics that quantifies the discrepancy between the predicted and measured images:

\begin{equation}
    \chi^2=\frac{1}{N_{\mathrm{DF}}} \sum_{i=1}^N\left(\frac{f^{(r)}\left(\nu_i\right)-f\left(\nu_i\right)}{\sigma_i}\right)^2,
\end{equation} where $N_{\mathrm{DF}}$ denotes the number of degrees of freedom, typically taken as the difference between the number of data points and model parameters. Iterations are terminated when $\Delta \chi^2$ of two successive iterations falls below a predefined threshold.

In the context of femtoscopy, the RL framework is applied by identifying the correlation function $ C(\mathbf{q}) $ as the observed image $f(\nu)$, the source function $S(\mathbf{r})$ as the target distribution  $\mathcal{F}(\mu)$, and the squared wave function $|\psi(\mathbf{q}, \mathbf{r})|^2 $ as the convolution kernel $ P(\nu|\mu) $. Using the experimentally measured correlation function and the theoretically computed wave function, the algorithm iteratively reconstructs the spatial source function.

\subsection{Establishment of Imaging Methods}

Based on the theoretical femtoscopy framework outlined previously, this section examines the effectiveness of imaging techniques for reconstructing source functions from correlation functions using test data.
We begin from the KP equation expressed in the out-side-long (o-s-l) coordinate system:

\begin{equation}
C(q_o, q_s, q_l) = \iiint S(r_o, r_s, r_l) , \left| \psi(\mathbf{q}, \mathbf{r}) \right|^2  \mathrm{d} r_o \mathrm{d} r_s \mathrm{d} r_l.
\end{equation}

To numerically implement the Richardson–Lucy algorithm, we discretize the continuous problem. The spatial domain in each of the three directions $(r_o,r_s,r_l)$ s divided into $N$ bins, and each momentum component $(q_o,q_s,q_l)$ is divided into $M$ bins. This discretization projects the three-dimensional correlation function onto the o-s-l directions.

The discretized correlation function data occupy $3M$ bins (across three momentum components), while the source function is defined over $N^3$ spatial voxels. To avoid an underdetermined system and reduce sensitivity to the initial guess, we maintain the condition $3 M > N^3$, which helps constrain the solution space and improves the convergence stability of the imaging algorithm.

In discrete form, the correlation function is approximated as:

\begin{equation}
C(q_o, q_s, q_l) = \sum_{i,j,k=1}^{N} K_{i,j,k}(q_o, q_s, q_l) S_{i,j,k},
\end{equation}where $i, j, k$ are spatial bin indices along the $r_o,r_s,r_l$ axes, $S_{i,j,k}$  denotes the discretized source function, and $K_{i,j,k}(q_o, q_s, q_l)$ is the discrete convolution kernel that incorporates the squared wave function and integration measure.

The kernel is constructed by averaging the squared wave function over the corresponding momentum and spatial bins:

\begin{equation}
\begin{split}
    K_{i,j,k}(q_o, q_s, q_l) = 
    \frac{1}{\Delta q^3} \int_{\mathrm{bin}(\mathbf{q})} d^3\mathbf{q}
    \left| \psi(\mathbf{q}, \mathbf{r}) \right|^2 \Delta r^3,
\end{split}
\end{equation}where $\Delta q$ is the momentum bin size and the spatial integration is performed over the volume of the $(i,j,k)^{\rm th}$ spatial bin. The momentum-space integration is evaluated via Monte Carlo sampling, which incorporates the experimental phase-space distribution and applicable momentum cuts. When computing a one-dimensional projection of the correlation function, appropriate limits are applied to the other two momentum components.

By systematically sampling the momentum bins and traversing all spatial voxels, one can compute the wave function and thereby construct the full convolution kernel required for the RL deconvolution procedure.

To validate the effectiveness of the RL algorithm for source function reconstruction, the model test is carried out according to the following procedure:

\textbf{Kernel Preparation}: The wave function convolution kernel is computed by sampling the momentum phase space, incorporating the relevant experimental acceptance and momentum cuts.

\textbf{Synthetic Data Generation}: A known source function is defined and convolved with the wave function kernel to obtain the true correlation function. Gaussian smearing is then applied to simulate experimental uncertainties and finite resolution.

\textbf{Imaging Process}: Starting with an initial source function estimate $SF^{(0)}(\mathbf{r})$, the RL algorithm is applied iteratively to the smeared correlation function. Convergence is monitored using the $\chi2$ statistic, defined as the relative error between correlation function predictions in successive iterations.

\textbf{Uncertainty Estimation}: A total of 10,000 independent correlation function datasets are generated and imaged. A single reconstruction is randomly selected as the central value, emulating the single‐sample nature of real experimental data. The standard deviation across all reconstructions is used to estimate the statistical uncertainty.

\section{Three-dimension Imaging Result}

The validation of the RL imaging algorithm begins with a controlled test using a Gaussian source function of known spatial form, as defined in Eq.~(3). This section details the parameter choices and presents the corresponding imaging results.

\begin{figure*}[hptb]
    \centering
    \includegraphics[width=0.9\textwidth]{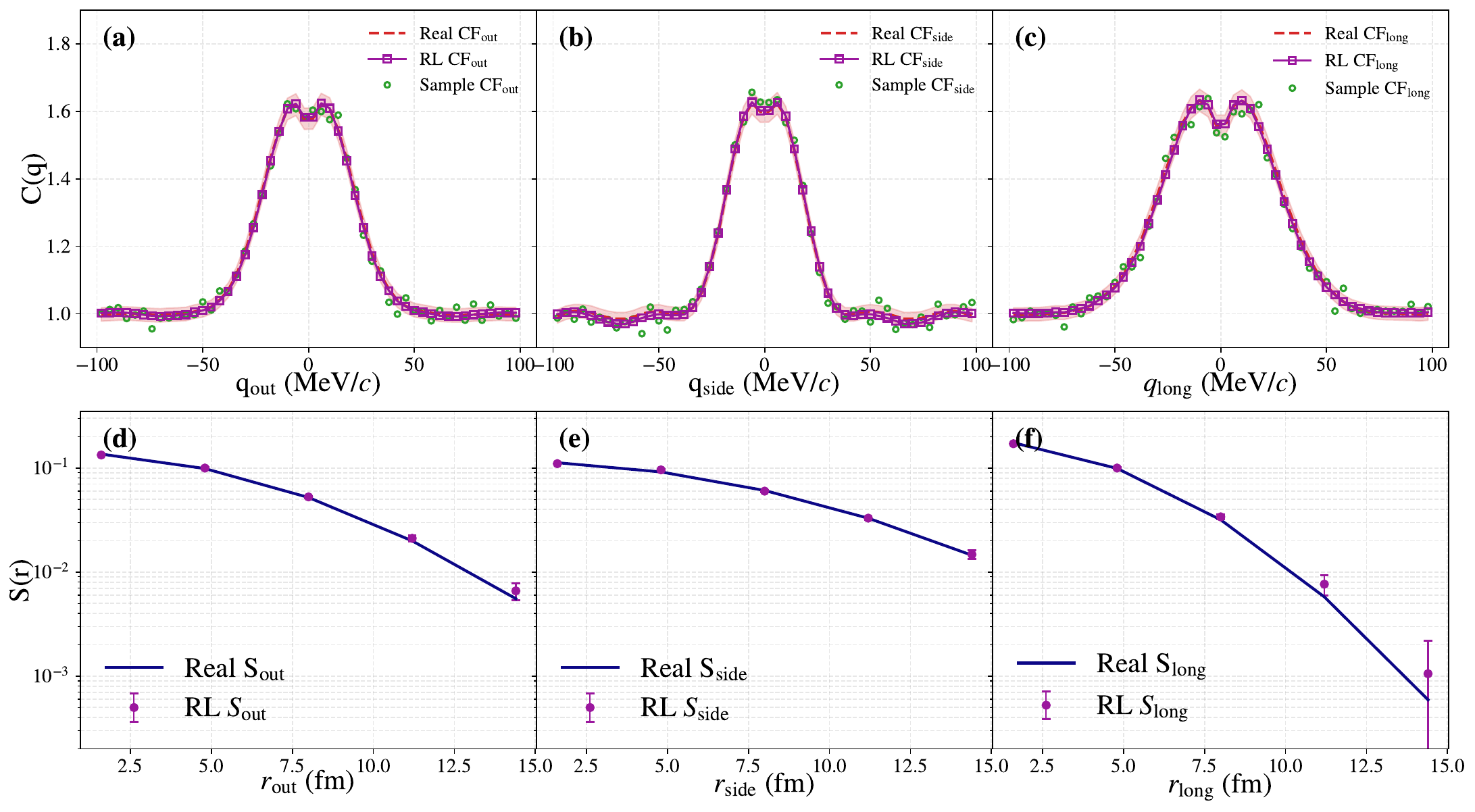}
    \caption{(a)--(c) Reconstructed correlation functions along the out, side, and long directions, compared with the true correlation functions. The restored results accurately reproduce the Coulomb-induced suppression at low relative momentum while reducing statistical noise. (d)--(e) Comparison between the reconstructed source functions and the true Gaussian distributions, showing excellent agreement in both shape and magnitude.}
    \label{fig2:gaus_CFSF}
\end{figure*}

\subsection{Gaussian Source Imaging}

A three-dimensional Gaussian source is adopted with true radii $R_o= 4 ~\mathrm{fm}, R_s= 5 ~\mathrm{fm}, R_l=3 ~\mathrm{fm} $. The spatial domain is divided into 5 bins over the interval $0 \text- 16~\mathrm{fm} $. The theoretical correlation function is obtained by convolving this source with the two-particle wave function. To mimic experimental data, Gaussian smearing with a width of $2\%$ of the true value is applied, which is used both as the “experimental” input and the starting point for RL iteration. It is referred to as the “sample correlation function” and is plotted as red circles in the panel (a)--(c) of Fig.~\ref{fig2:gaus_CFSF}. The momentum phase space is modeled as a three-dimensional Gaussian with a width of $ 0.2~\mathrm{GeV}/c$  in each direction. In the Monte Carlo sampling, the transverse momentum $k_{\rm T}$ is restricted to $0.2-0.25 ~\mathrm{GeV}/c$, and when sampling one momentum component, the other two are limited to less than $12~\mathrm{MeV}$, consistent with typical experimental selections.

The initial source function for the RL iteration is chosen as $R_o = R_s = R_l = 4.5 ~\mathrm{fm}$, and the convergence threshold is set at $\Delta\chi^2<10^{-6}$, where $\Delta\chi^2$ is the difference of $\chi^2$ between two successive iterations, and $\chi^2$ is defined as the relative difference between correlation functions from consecutive iterations. Tests with different initial values confirm that the imaging result is robust against this choice.

As shown in panels (a)--(c) of Fig.~\ref{fig2:gaus_CFSF}, the reconstructed correlation functions along the out, side and long directions agree closely with the true distributions. The restored correlation function closely matches the true underlying curve, successfully reproducing the central depression caused by Coulomb repulsion while reducing statistical fluctuations.

Panels (d)--(e) present the reconstructed source functions alongside the true Gaussian distributions. For computational efficiency and clarity, only the one-sided profiles are shown, leveraging the reflection symmetry inherent in each direction. The imaged source functions demonstrate remarkable consistency with the true spatial distributions, validating the accuracy of the reconstruction method.

\begin{figure}[hptb]
    \centering
    \includegraphics[width=0.5\textwidth]{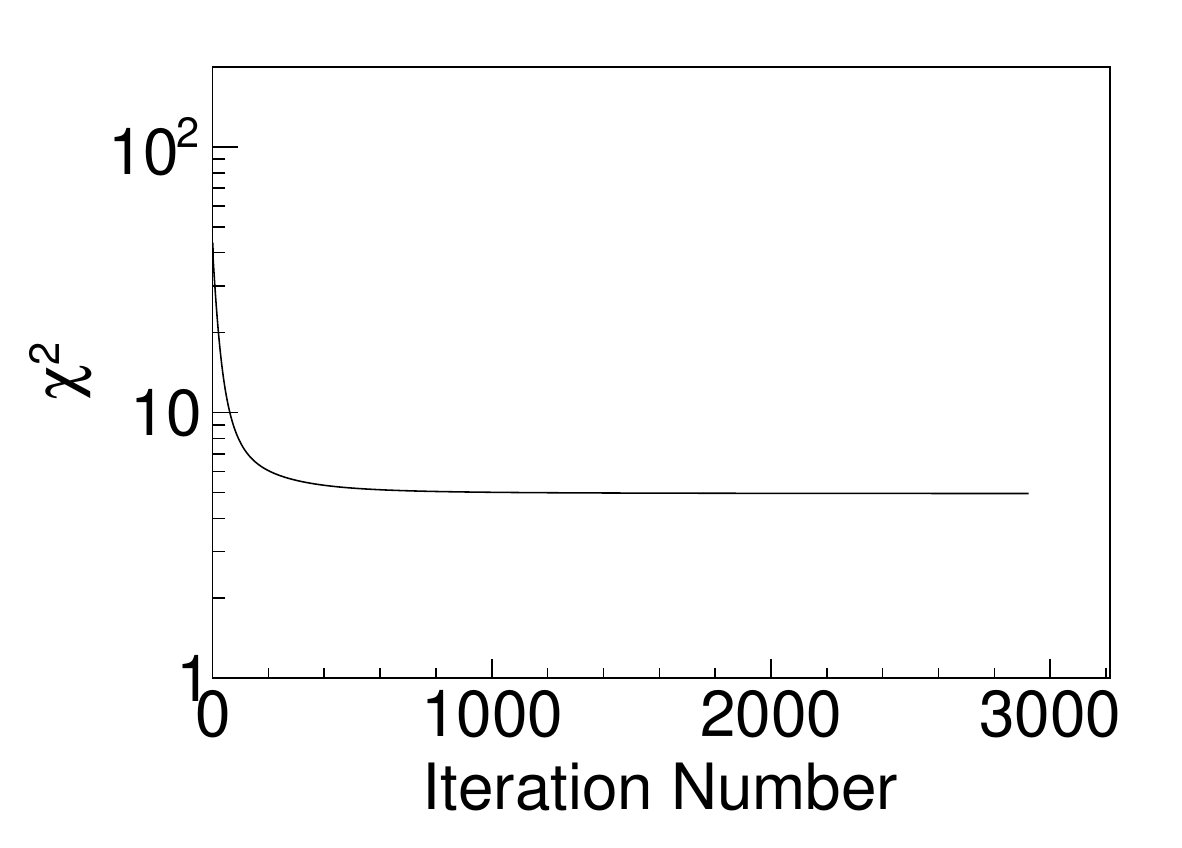}
    \caption{Evolution of $\chi^2$ with the number of iterations}
    \label{fig:chi2}
\end{figure}

\begin{figure}[hptb]
    \centering
    \includegraphics[width=0.5\textwidth]{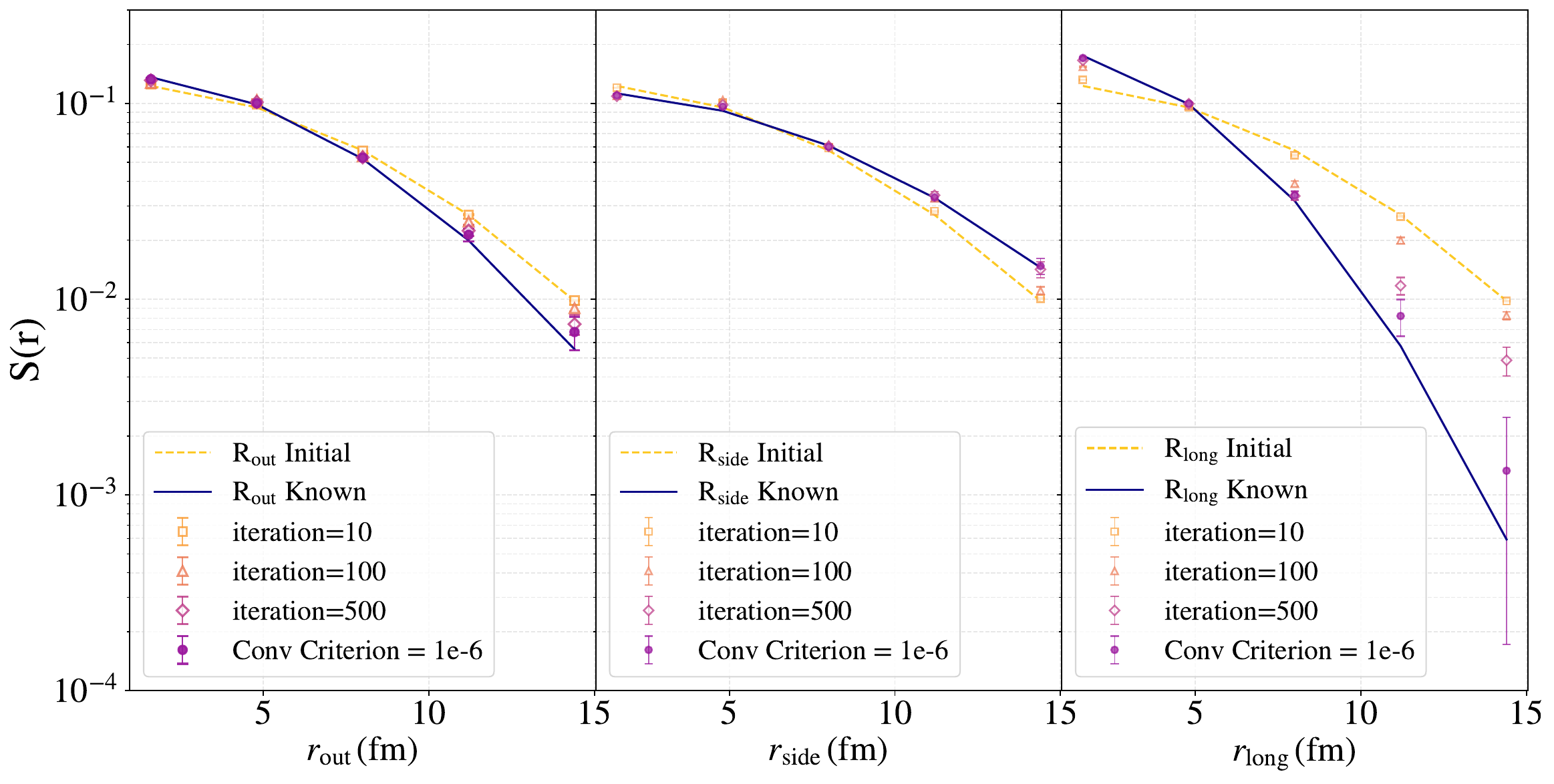}
            \caption{Convergence of the imaged source function. Open symbols are  the reconstructed profile at $10^{\rm th}$,  $100^{\rm th}$ and  $500^{\rm th}$ iteration, evolving from the initial guess (dashed) toward the final deblurred distribution (filled circles), which overlaps with the input known distribution (solid).}
    \label{fig:iter}
\end{figure}

\begin{figure*}[hbt]
    \centering
    \includegraphics[width=0.9 \textwidth]{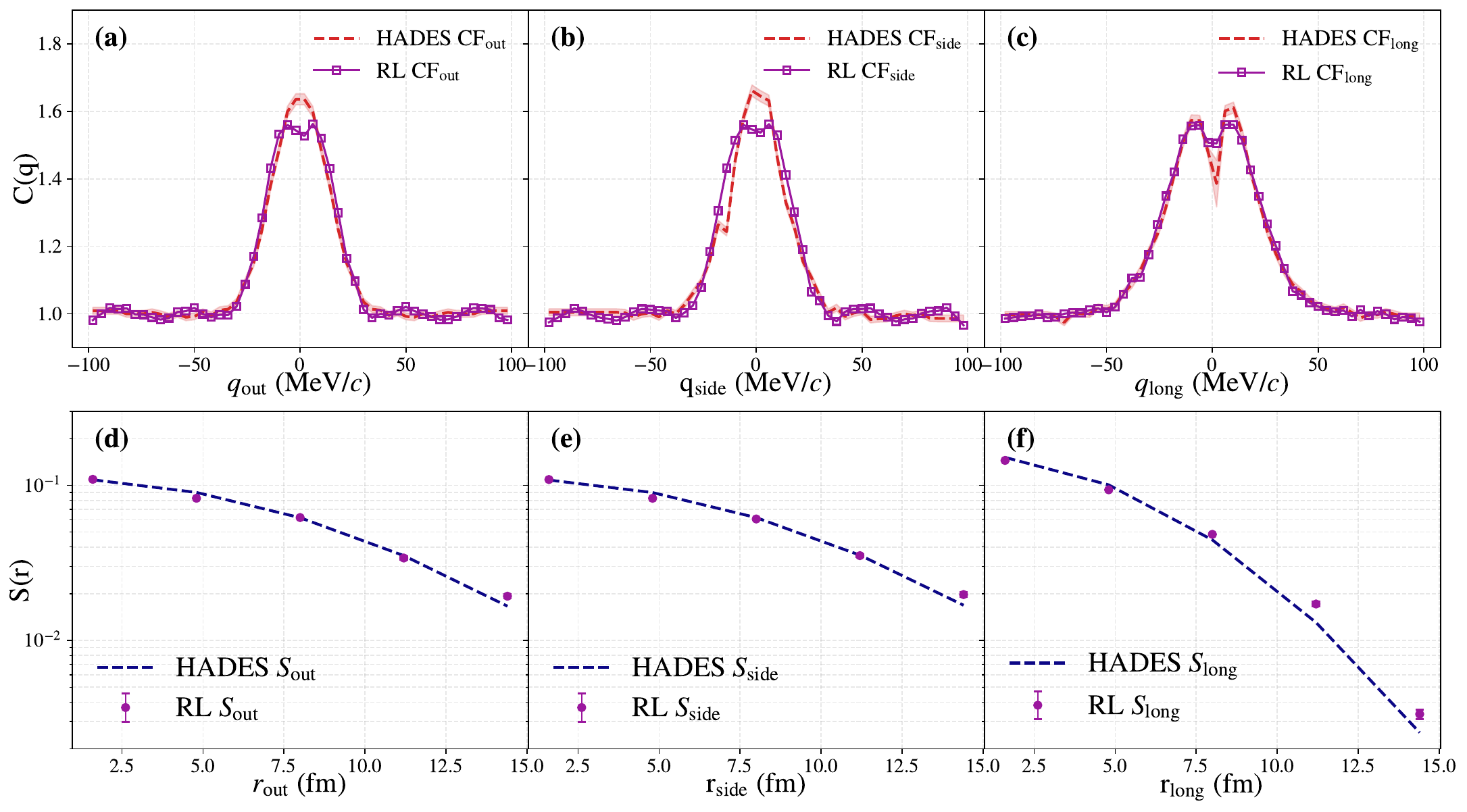}
    \caption{(a)--(c) Reconstructed correlation functions along the out, side, and long directions, compared with the experimental data from HADES. The light red bands indicate experimental uncertainties. Slight discrepancies in the central region may result from the algorithm's treatment of Coulomb repulsion effects. (d)--(f) Comparison between the imaged source functions and the Gaussian-parameterized source functions from HADES. The reconstructed profiles show an upward deviation at large relative distances $r$ particularly on the longitudinal side, suggesting non-Gaussian characteristics in the emission source.}
    \label{fig:hades}
\end{figure*}

The convergence behavior of the RL algorithm is illustrated in Figs.~\ref{fig:chi2} and \ref{fig:iter}. Figure~\ref{fig:chi2} displays the evolution of the $\chi^2$ as a function of iteration number The convergence criterion is based on the rate of change of $\chi^2$ between successive iterations. The $\chi^2$ value decreases rapidly and stabilizes, indicating stable convergence without significant overfitting—a known concern in RL reconstructions when iterations continue excessively \cite{liuNoiseAmplificationIllconvergence2025a}. In our implementation, the limited number of imaging points and small uncertainties help mitigate such potential issues.

Figure~\ref{fig:iter} illustrates the corresponding progression of the reconstructed source function with increasing iteration count.
As iterations proceed, the imaged profile systematically approaches the true spatial distribution. A criterion of $10^{-6}$ proves appropriate in this setting, balancing reconstruction accuracy and computational cost.

These results confirm that the RL algorithm performs with high accuracy in three-dimensional source imaging. In the following section, the method is applied to real experimental data from the HADES Collaboration to extract the source function, which will be compared with the conventional Gaussian-parameterized form.

\subsection{Experimental Data Imaging}

This study analyzes experimental data from the High Acceptance Di-Electron Spectrometer (HADES) at the SIS18 heavy-ion synchrotron of GSI, Darmstadt \cite{HADES:2009aat,HADES:2018gop}. The data were collected from central Au+Au collisions at a beam energy of $1.23~\mathrm{AGeV}$, with identical pion pairs used for intensity interferometry to access femtoscopic properties of the collision zone.


In the analysis, pions are selected within a rapidity window of $|y - y_{\rm cm}|<0.35$ and a transverse momentum interval of $0.2-0.25~\mathrm{GeV}/c$. The initial source radii used in our imaging procedure are taken directly from the Gaussian parameterization given from the paper\cite{HADES:2019lek}: $R_o = 5.01 ~\mathrm{fm}, R_s = 4.94 ~\mathrm{fm} $, and $R_l = 3.38~\mathrm{fm}$. These values serve as the starting point for the RL iteration to facilitate convergence.

Panels (a)–(c) of Fig.~\ref{fig:hades} compare the experimentally measured correlation function with the RL-reconstructed result. An obvious depression is observed at low relative momentum $\mathbf{q}$, arising from Coulomb repulsion encoded in the wave function. The deviations at larger momenta may be attributed to the use of wide spatial bins in the source function imaging, which limits the resolution of finer structural details in the emission distribution.

Panels (d)--(f) of Fig.~\ref{fig:hades} display the source function obtained via the Sinyukov parameterization—used as the initial input—together with the final imaged source function. A notable difference emerges at large relative distances $\mathbf{r}$, where the imaged source exhibits a clear upward deviation. This enhancement beyond the uncertainty likely reflects non-Gaussian features of the emission source, which has also been observed in the imaging result at RHIC energies \cite{Xu:2024dnd,Kincses2025}, and suggestively corresponds to physical origin that the colliding system is not fully randomized at the freeze-out instant.

\section{Probing Neutron Skin Effects via Femoscopical Imaging}

This section investigates the sensitivity of $\pi\pi$ correlation observables to the neutron skin thickness in colliding nuclei using the Ultra-relativistic Quantum Molecular Dynamics (UrQMD) model combined with the RL imaging algorithm. Simulations are performed for Pb+Pb collisions at 1.5 AGeV, a system widely used in neutron skin studies due to its theoretical relevance and the high yield of $\pi^-$ mesons, which enables statistically robust femtoscopic measurements. Unlike recent studies  on Gaussian HBT radii \cite{Li:2025mox}, our imaging approach reconstructs the 3-dimensional spatial structure of the source, offering a more detailed view of neutron skin effects.


\subsection{Framework Description}

The Ultra-relativistic Quantum Molecular Dynamics (UrQMD) model is a microscopic transport approach that simulates the dynamics of heavy-ion collisions across a broad energy range. It describes the evolution of hadrons via covariant equations of motion, incorporating binary collisions, resonance formations, and decays, along with a range of interaction potentials.

In UrQMD, each hadron is represented as a Gaussian wave packet. The system evolves according to Hamilton’s equations:
\begin{equation}
    \dot{\boldsymbol{r}}_i = \frac{\partial H}{\partial \boldsymbol{p}_i}, \quad \dot{\boldsymbol{p}}_i = -\frac{\partial H}{\partial \boldsymbol{r}_i},
\end{equation}
where the Hamiltonian $ H_i = E_i^{\text{kin}} + U_i $  includes a potential term composed of Skyrme (two- and three-body) and Coulomb interactions:
\begin{equation}
V = V^{(2)}_{\text{Sky.}} + V^{(3)}_{\text{Sky.}} + V_{\text{Coul.}}
\end{equation}
The single-particle potential $U$ is derived from the functional derivative $U=\delta V / \delta f$ of the total interaction potential with respect to the phase-space distribution $f(\mathbf{r}, \mathbf{p})$.

To probe neutron skin effects, the initial neutron density distribution in Pb nuclei is modeled using a droplet-model form with a Woods–Saxon potential:
\begin{equation}
\rho_i(r)=\frac{\rho_i^0}{1+\exp\left(\frac{r-R_i\left[1-\left(0.413 f_i t_i / R_i\right)^2\right]}{f_i t_i / 4.4}\right)}, \quad i=\text{n, p},
\end{equation}
where $t_i = 2.18\ \mathrm{fm}$ is the diffuseness parameter, and $f_i$ controls its scaling. We fix $f_{\rm p} = 1.0$ for protons and vary $f_{\rm n}$ to adjust the neutron skin thickness $\Delta R_{\rm np}$. Two configurations are calculated and compared with $f_{\rm n} = 1$ and $f_{\rm n} = 6$, corresponding to different neutron density profiles. For each, the correlation functions are calculated, followed by 3-dimensional imaging of the source functions using RL algorithm.

\begin{figure}[hptb]
    \centering
    \includegraphics[width=0.45\textwidth]{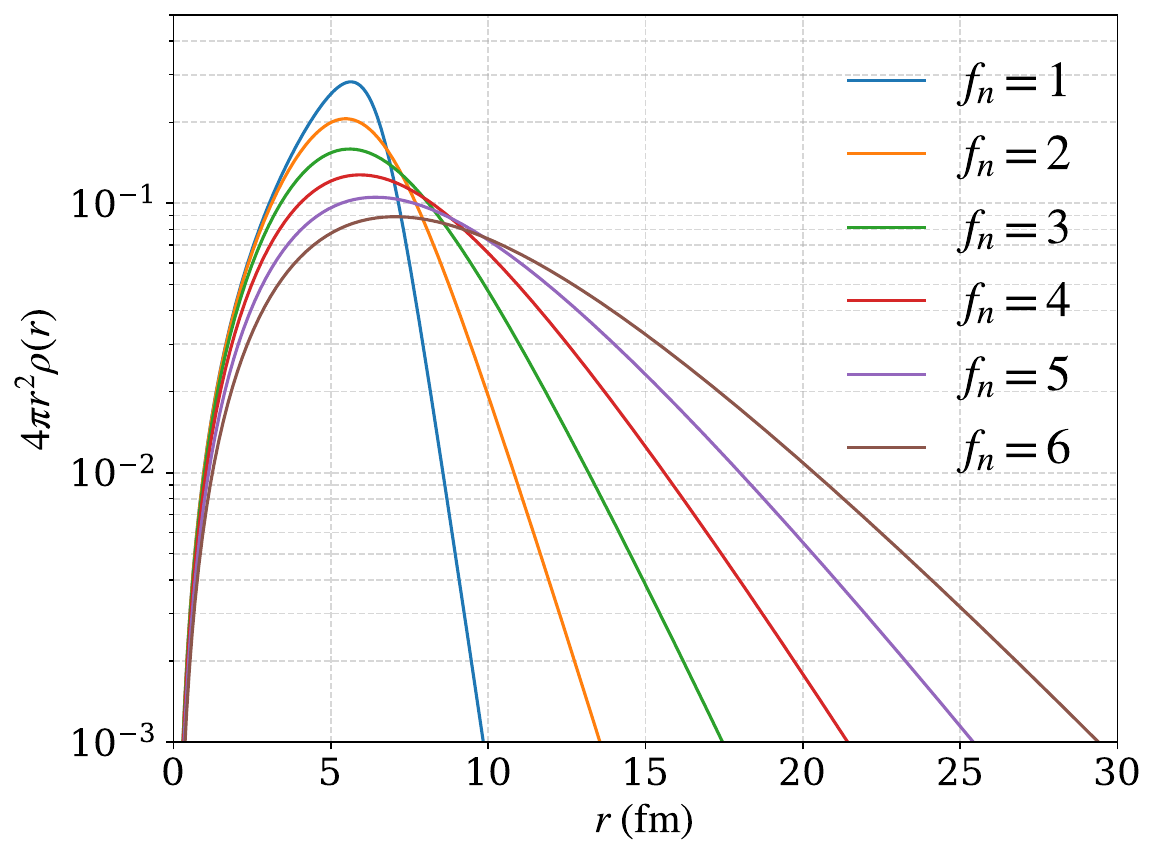}
    \caption{ Neutron probability distribution $4\pi r^2 \rho_{\rm n}(r)$ of Pb for different values of the diffuseness scaling parameter  $f_{\rm n}$ (see text). }
    \label{figrho}
\end{figure}

Figure~\ref{figrho} illustrates the neutron density of Pb profiles obtained from Eq.~(17) as a function of the radial coordinate $r$ for $f_{\rm n}$ values ranging from 1 to 6. As $f_{\rm n}$increases, the neutron distribution becomes increasingly diffuse, demonstrating that $f_{\rm n}$ serves as an effective parameter for varying the neutron skin thickness. Our simulations aim to determine whether the change of  $f_{\rm n}$ brings  discernible effects in the correlation functions and the deblurred source function.

\begin{figure*}[hptb]
    \centering
    \includegraphics[width =0.9\textwidth]{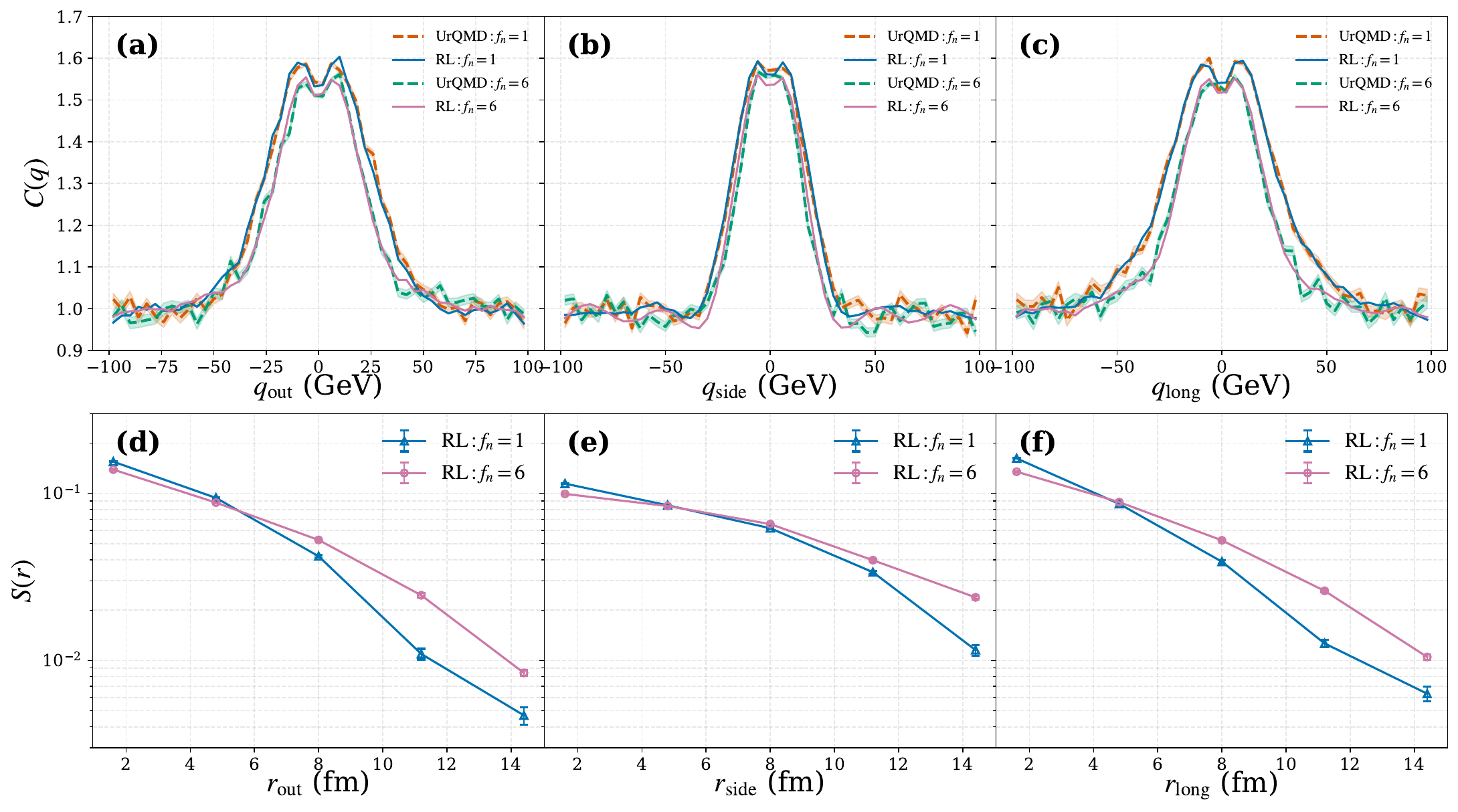}
    \caption{ Correlation functions and reconstructed source functions of $\pi^-\text- \pi^-$ pair in 1.5 AGeV Pb+Pb collisions.  Upper: Three dimensional correlation functions  along out (a), side (b) and longitudinal (c) directions with neutron skin parameter $f_{\rm n}=1$ and $f_{\rm n}=6$ for lead. Both the simulation result of UrQMD model (dashed) and the RL reconstructed (solid) are presented. Lower: The corresponding source functions extracted by RL algorithm with  $f_{\rm n}=1$ and $f_{\rm n}=6$.}
    \label{figurqmd}
\end{figure*}

All simulations use UrQMD v3.0.5. Further details are available in \cite{Li:2008bk, Li:2022iil, Hillmann:2018nmd, Hillmann:2019wlt, Steinheimer:2022gqb}, and neutron density parametrizations are described in \cite{Li:2025mox, Sun:2009wf}.
 
\subsection{Sensitivity of the Source Function on Neutron Skin Thickness}

We simulate central Pb+Pb collisions at 1.5 AGeV with zero impact parameter  ($b = 0$ fm) using the UrQMD model. The model provides complete phase-space information, including momenta and spatial coordinates, of all emitted particles at freeze-out. For each event, we analyze all possible $\pi^-\pi^-$ pairs and compute their relative momentum components in the o-s-l coordinate system. The squared modulus of the two-particle Coulomb wave function is used as a weighting factor to fill three-dimensional histograms of the correlation function. When projecting along one momentum direction, the other two components are constrained to magnitudes below 12 MeV. Pairs are further selected within the transverse momentum range $k_{\rm T} = 0.2–0.25 ~\mathrm{GeV}/c$ .

A total of $3\times10^6$ events are generated for each neutron skin configuration. For clearness, the resulting correlation functions for $f_{\mathrm{n}}=1$ and $f_{\mathrm{n}}=6$ are presented as dashed curves in panels (a)–(c) of Fig.~\ref{figurqmd}, with light-colored bands indicating statistical uncertainties. The central values of these distributions are used as the input to the RL deblurring algorithm, and the reconstructed correlation functions are shown as solid curves, which again reproduces well the initial ones, demonstrating the robustness of the approach. It is also seen that the correlation function for $f_{\mathrm{n}}=1$ exhibits a broader distribution compared to that for $f_{\mathrm{n}}=6$.

Panels (d)–(f) display the corresponding reconstructed source functions. The source function for the larger $f_{\mathrm{n}}$ value (thicker neutron skin) shows a more diffuse spatial distribution, in accordance  with the increase in HBT radii reported in Ref.~\cite{Li:2025mox}, particularly in the out and side directions. This systematic broadening demonstrates the sensitivity of the imaged source function to variations in the neutron density distribution, validating RL deblurring algorithm  as a promising tool for probing nuclear structure through heavy-ion collisions. The dependence of the HBT radii on the impact parameter has been investigated in Ref.~\cite{Li:2025mox}, where a Gaussian source function is assumed.

\section{Summary}


To summarize, the Richardson–Lucy deconvolution algorithm has been extended to reconstruct three-dimensional pion source functions from two-particle correlation measurements. The method was tested using simulated data from Gaussian sources, demonstrating its stability with respect to initial conditions and its ability to recover the underlying spatial distribution with promising accuracy. When applied to experimental data from HADES Au+Au collisions at 1.23 AGeV, the approach proved effective under realistic conditions, revealing non-Gaussian features in the extracted source function, particularly at large relative distances.

Additionally, the RL imaging method has been applied to examine the sensitivity of the reconstructed source function to the neutron skin thickness parameter of lead based on UrQMD simulations of Pb+Pb collisions at 1.5 AGeV. By varying the neutron density profile within the droplet model, it is observed that the imaged source function responds distinctively to changes in the neutron skin configuration. This result underscores the potential of correlation function imaging as a direct probe of nuclear density distributions, offering a complementary perspective to traditional Gaussian HBT parameterizations.

The Richardson–Lucy algorithm provides a robust, model-independent tool for 3-dimensional femtoscopic imaging of the particle emission source. The sensitivity of the imaged source to the neutron skin thickness opens new pathways for exploring nuclear structure properties in heavy-ion collisions. Eventually, systematic applications to high-quality datasets of correlation functions  promise to deepen our understanding of collision dynamics and the equation of state of nuclear matter.

\textbf{{Acknowledgement}} This work is supported by the National Natural Science Foundation of China under Grant Nos. 12335008 and 12205160, by the Ministry of Science and Technology under Grant No. 2022YFE0103400,  by the Center for High Performance Computing and Initiative Scientific Research Program in Tsinghua University.

\bibliography{reference}

\end{document}